\documentstyle[aps,epsfig]{revtex}

\newcommand{\be}{\begin{equation}}
\newcommand{\ee}{\end{equation}}
\newcommand{\up}{\uparrow}
\newcommand{\dn}{\downarrow}
\renewcommand{\vec}[1]{{\bf #1}}

\begin{document}


%
\title{Full-CI Quantum Chemistry using the Density-Matrix Renormalization Group}
\author{S. Daul $^1$, I. Ciofini $^2$, C. Daul $^2$ and Steven R. White $^3$}
\address{ 
 $^1$ Institute for Theoretical Physics, University of California, \\
              Santa Barbara CA 93106. \\
 $^2$ Institut de Chimie Inorganique et Analytique, Universit\'e de Fribourg,\\
	    CH-1700 Fribourg, Switzerland. \\
 $^3$ Department of Physics and Astronomy, University of California, \\
	Irvine CA 92697.
}

\maketitle

\begin{abstract}

We describe how density-matrix renormalization group (DMRG)
can be used to solve the full-CI problem in quantum chemistry.
As an illustration of the potential of this method, we apply it to a 
paramagnetic molecule. 
In particular, we show the effect of various basis set, the scaling
as the fourth power of the size of the problem, and compare the DMRG with
other methods.

\end{abstract}

\section{Introduction}

First-principle quantum chemistry is employed successfully to
obtain thermochemical data; molecular structures; force fields and
frequencies; assignments of NMR-, photoelectron-, E.S.R-, and UV-spectra;
transition state structures as well as activation barriers; dipole
moments and other one- or two-electron properties. 
Two routes of calculation are available:
(i) ab initio Hartree-Fock (HF) and post-HF methods; and
(ii) density functional theory (DFT). 
Though both approaches are rigorous, the former
one necessitates lengthy configuration interaction (CI) treatment
to account for electron correlation; whereas the latter one crucially
depends upon the quest for accurate exchange and correlation functionals.
The recently acquired popularity of DFT stems in large measure from its
computational efficiency, allowing it to treat medium to large size 
molecules at a fraction of the time required for HF or post-HF
calculations.  
More importantly, expectation values
derived from DFT are, in most cases, better in line with
experiment than results obtained from HF calculations. 
This is particularly the case for systems involving transition metals. 
Nevertheless, if one wants to 
achieve experimental accuracy for small polyatomic molecules,
the method reaches its limits. 
On the other hand, post-HF overcomes these limits and goes even 
beyond experimental accuracy, e.g. in case of H$_2$. 
The drawback is of course its very high cost in computational power. 
However these very accurate calculations of small systems may 
provide the best route to obtaining more accurate 
exchange-correlation potentials using the 
constrained search 
algorithm of Levy \cite{Levy}.

Keeping the discussion to ab initio (HF and post-HF) quantum
chemistry, let us now consider the state of the art in this topic. 
We now know how to do very large calculations, using
the direct methodology \cite{Almlof}. 
We can also manage to work with good basis sets for such calculations, 
although it is
considered that 6-31G* are not good enough, and probably something nearer to
TZ2P is required for definitive SCF calculations.
Beyond SCF there are major difficulties, all associated with
trying to more accurately represent the electron-electron cusp. We know from
the work of Kutzelnigg \cite{Kutz}, 
that the convergence of this problem is very 
slow, something like $(\ell+\frac{1}{2})^{-4}$, with $\ell$  the
orbital angular momentum quantum number. 
This means that very large basis sets are required for correlated 
calculations. 
We know that it is more important to
include $d$ and $f$ basis functions than to improve the methodology
(e.g. 6-31G* basis are not appropriate for correlated studies). 
We also know that the raw cost of the main correlated methods, 
MP2, MP3, MP4, CISD, QCISD, and QCISD(T)
increase with powers of the size of the problem as 5, 6, 7, 6, 6, 7,
respectively.
We recognize that QCISD(T), with large basis sets, can make predictions 
for small molecules which are of ``chemical accuracy''
(e.g. the G2 quantum chemistry of J. Pople at al.).
Considering ``coupled cluster" technology, although this method is very 
promising, we have to realize that it is not yet possible to contemplate 
calculations with 1000 basis functions.
The rapid
progress that computer companies have made in the development of their
hardware, such as more memory, vectorization compilers, and parallel
machines will not cure these basis set problems.

Recently, one of us (S.W.) and R. L. Martin have shown that
with the density matrix renormalization group (DMRG), one can calculate
``exactly" (full CI) the electronic structure of N interacting electrons in a
field of M nuclei \cite{WhiteMartin}. 
By ``exactly" we mean that the correctness of the
treatment of correlations increases very rapidly and systematically
with the calculational effort, so that one is essentially
only limited by the accuracy of representing the Hamiltonian in 
a finite basis set. 
Moreover, these authors showed that the method theoretically scales as
${\cal O}(N^4)$ and hence the calculation of much larger molecules will be
feasible.
More precisely, this $N^4$ dependence originates from the summation over 
four indices in the second quantized form of our Hamiltonian (Eq. 13). 
Hence, it will be possible to exploit the short sightedness of
interatomic interactions in the future e.g. in using localized basis function 
(cf. vide infra).
Thus we expect to achieve linear scaling, ${\cal O}(N)$, for large linear 
molecules when teraflop CPU's and
peta bytes storage media will become routinely available to us.
Moreover, we would like to point out that the method at hand is well suited for 
CAS calculations. 
Indeed, since the basis functions used will usually be Hartree-Fock or 
Kohn-Sham eigenfunctions, we can easily use the
corresponding eigenvalues to setup an active space by discarding 
 inner shell orbitals (in keeping their occupation frozen) 
or in eliminating those with very large energies. 

This paper is organized as follows. In Sec. II, we recall the physical
principles of the DMRG. Then we reformulated, in a more chemical language, 
the basic algorithm for molecules in Sec. III.
Finally, in Sec. IV we apply the technique on a paramagnetic molecule and
discuss the performance of the method as a function of the basis set used.

\section{Principles of The Density Matrix Renormalization Group (DMRG)}

\subsection{The Renormalization Group}

The idea of the Renormalization Group (RG) is to apply a transformation to the 
Hamiltonian which eliminates degrees of freedom that are unimportant for
the description of the states in the energy range we are interested in.
For example, if we are interested in the low energy states of a system
with a energy cutoff $\Lambda$, one integrates out modes with energy
\begin{equation}
        \Lambda-\delta \Lambda \leq \omega \leq \Lambda 
\end{equation}
where $\delta \Lambda$ is a small energy interval. 
Then we rescale the parameters of the new system so that it reproduces
the previous one. 
In other words, given a Hamiltonian $H$ of a system  with $N$ variables,
a renormalization transformation $R_b$ is a mapping in the Hamiltonian
space which maps $H$ to $H'$,
\begin{equation}
  H' = R_b(H).
\end{equation}
This new Hamiltonian describes a system with $N' = N/b \leq N$ variables.
The RG transformation must be unitary, i.e. it has to preserve the partition
function ($Z = \mbox{Tr}\; e^{-H/kT} $),
\begin{equation}
       Z_{N'} \left[ H'  \right] = Z_{N} \left[ H  \right].
\end{equation}
This set of transformations is called a group because the transformations
must be associative:
\begin{equation}
        R_{b'} \left( R_b(H) \right) =  R_{b'b}(H).
\end{equation}
An exact transformation is usually not possible. 
For example, Wilson and Fisher  worked in 
Fourier space and used a perturbative scheme in order to analytically solve
this problem \cite{WilsonKogut}. 
Another approach is to consider the Hamiltonian on a lattice and to work
in real space. 
A RG transformation could be the elimination of every second
site and then the redefinition of the Hamiltonian parameters in 
order to reproduce the original problem.

A numerical implementation for a quantum mechanical problem was given
by Wilson who succeeded in constructing a non-perturbative RG transformation 
for the Kondo model \cite{Wilson}.
The same transformation has since then been applied to a wide range of
similar quantum impurity models with equal success \cite{Krishnamurthy}.
On the other hand, it has proven to be more difficult to develop similar RG 
transformation for quantum lattice models.
These difficulties with RG approaches in real space provided the motivation
for the development of DMRG.

\subsection{Wilson's RG}

Let $H$ be a one-dimensional Hamiltonian describing an interacting system
of electrons on a lattice with $L$ sites labeled $i$. 
Each sites has one (Wannier) orbital which can be in 4 different
states : $ | 0 \rangle$, $ | \uparrow \rangle$, $| \downarrow \rangle$ 
 and $ | \uparrow \downarrow \rangle  $.
The dimension of the Hilbert space for $N_\uparrow$ up-spin electrons and
$N_\downarrow$ down-spin electrons is thus
\begin{equation}
\mbox{dim} {\cal H} =
\left(  \begin{array}{c} 
	L \\ 
	N_\uparrow 
  \end{array} \right)
\left(  \begin{array}{c} 
	L \\ 
	N_\downarrow
  \end{array} \right).
\end{equation}
Which is, for $L=100$ and  $N_\uparrow = N_\downarrow =50$, 
$\mbox{dim} {\cal H} = 10^{58}$.
This is of course intractable numerically. 
The idea is to obtain the low energy eigenstates of that system by 
keeping only a small number of states, e.g. 1000.
	
For the sake of clarity we will first directly apply  Wilson's scheme to quantum 
lattice systems and then show why it does not work with a simple example.
We will describe next a RG step. 

\begin{figure}[hbt]
\begin{center}
  \epsfig{file=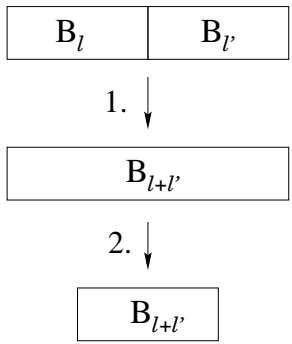,height=45mm}
\end{center}
\caption{Wilson's blocking scheme for real-space renormalization group.}
\label{fig:stdrg}
\end{figure}

Let $B_\ell$ be a block describing the first $\ell$ sites for which we 
only keep $m$ states to describe the Hamiltonian of that part of the system.
The internal Hamiltonian $H_\ell$ is then represented by 
an $m \times m$ matrix. 
One adds to this block (see Fig. \ref{fig:stdrg}) another block  $B_{\ell'}$
describing $\ell'$ sites with internal Hamiltonian $H_{\ell'}$.
The Hamiltonian of the new block $B_{\ell+\ell'}$ has dimension
$m m'$ and can be written as
\begin{equation}
  H_{\ell+\ell'} = H_{\ell} \otimes {\bf 1} +  {\bf 1} \otimes H_{\ell'}
+ \sum_\alpha c_\alpha A^\alpha \otimes B^\alpha .
\end{equation}
Here, the last term represents the interactions between $B_\ell$
and $B_{\ell'}$, where the $A^\alpha$ are operators in $B_\ell$
and the $B^\alpha$ are in $B_{\ell'}$ (See Section III.C.)
The RG idea is then to reduce the size of this many body basis
from $m m'$ to $m $. 
For this purpose, in Wilson's original formulation,  one diagonalizes 
$ H_{\ell+\ell'}$ and keeps only the $m$ 
states with the lowest energy eigenvalues, since we seek the low energy states.

\begin{figure}[hbt]
\begin{center}
  \epsfig{file=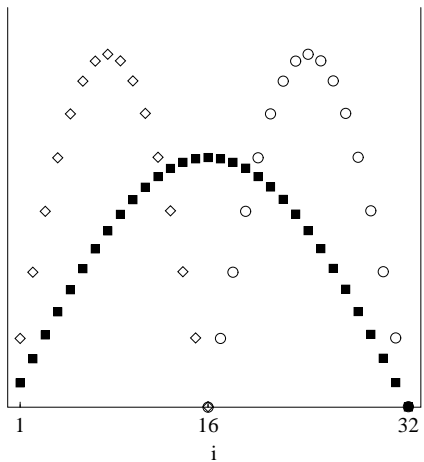,width=55mm}
\end{center}
\caption{The lowest eigenstates of two 16-sites system (open circles) and 32-sites 
system (full squares) for a one-dimensional free particle with open boundary 
conditions. }
\label{fig:pbox}
\end{figure}

This method has been directly applied to quantum lattice systems \cite{Chui}
but the spectra become inaccurate after a few iterations. 
The problem originates from the treatment of the boundary conditions.
To illustrate this, we will consider the problem of a free particle on a 
lattice of 
size $L$ with open boundary conditions or, in other words, the discretized
version of a free particle in a box. 
The eigenstates of this system are
\begin{equation}
 \chi_n(x) \sim \sin \left( \frac{n \pi }{L} x \right) 
\end{equation}
where $n$ is a positive integer.
We now consider the addition of two blocks of size $L$. 
In the RG procedure, the lowest few eigenstates of system of length $L$
are combined to form the low-lying eigenstates of the system of length $2L$.
The lowest eigenstates of two systems of length $L$ and of a single
system of length $2L$ are shown in
Fig. \ref{fig:pbox}. 
One sees clearly that a combination of the ground states of the two small
systems (size $L$) cannot reproduce the ground state of the large system 
(size $2L$).
The kink in the superposition of the two ground states of the small systems
can only be removed if one keeps almost all the eigenstates of the smaller 
systems.

This example shows that the treatment of the boundaries of the block is 
crucial.
White and Noack \cite{WhiteNoack92} formulated two types of RG procedures
which solve this problem. 
Both are based on choosing a new basis for  $H_{2L}$ which is not the
set of eigenstates of   $H_{L}$.
In their first procedure, the combination of boundary conditions method, the new
basis is obtained from the low-lying eigenstates of several different block
with different boundary conditions at the edge of the block. 
In the second type of procedure they proposed, 
the superblock method, the new basis for $H_{2L}$ exploits the idea that it
will eventually be used to make up parts of a larger system. 
These two methods were applied to the free particle in a box and both yielded
accurate results. 
While the first method is not so well suited for many particle problems, the 
second one necessitates an optimal projection from the superblock.
This gave rise to the DMRG scheme which is described in the next section.

\subsection{Density-Matrix approach}

The fundamental difficulty in the standard approach lies in the choice of 
the lowest energy eigenstates of $H_{\ell + \ell'}$ as the new truncated basis. 
Since $H_{\ell + \ell'}$  contains no connections to the rest of the lattice, 
its eigenstates exhibit inappropriate features at the block boundaries. 
One way of avoiding this problem is to diagonalize a larger system 
(a ``superblock'')  which contains the new block $B_{\ell + \ell'}$. 
The idea is that fluctuations in the rest of the superblock 
(the ``environment'') lead to a more appropriate  treatment of the boundary 
of  $B_{\ell + \ell'}$.
The wave function for the superblock is projected onto $B_{\ell + \ell'}$
producing a set of states of $B_{\ell + \ell'}$ which are then retained.

Let $\{|i\rangle\}$ be a complete set of states of  $B_{\ell + \ell'}$ 
and $\{|j\rangle\}$ be the states of the rest of the superblock. 
We can then write 
\be
   |\psi \rangle = \sum_{i,j} \psi_{ij}|i\rangle |j\rangle .
\ee
In Ref. \onlinecite{White92} one of us (S.W) shows that the optimal states to be kept 
are the 
eigenvectors of the reduced density matrix of the superblock, 
\be
               \rho_{ij} = \sum_k \psi_{ik}\psi_{jk} .
\ee
Let us denote by $\omega_{\nu}$ the eigenvalues of $\rho$ in decreasing order.
The $m$ optimal states we seek,
$u^{\nu}$, are the eigenstates of $\rho$ with the largest eigenvalues. 
Each $\omega_{\nu}$ represents the probability of the system being in the 
state $u^{\nu}$, hence $\sum_{\nu} \omega_{\nu} =1$. 
The deviation of 
\be
    P_m = \sum_{\nu=1}^m \omega_{\nu} 
\ee
from unity measures 
the accuracy of the truncation of the $m$ states (typically, $1-P_m \leq 10^{-6}$).

\subsubsection{Density-Matrix algorithm}

\begin{figure}[hbt]
\begin{center}
  \epsfig{file=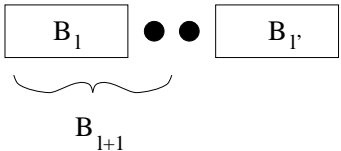,height=2cm}
\end{center}
\caption{The configuration of blocks used for the density matrix calculations.}
\label{fig_dmrg_block}
\end{figure}

In practice we take one block to be just a single site. 
Fig. \ref{fig_dmrg_block} shows the superblock configuration 
that we use. 
We adopt the notation $B_\ell \bullet \bullet B_{\ell'}$ for this configuration,
where $B_\ell$ represents a block composed of $\ell$ sites, $B_{\ell'}$ is a
block of length $\ell'$, $\bullet$ represents a single site, and the length of
the superblock is therefore $L=\ell +\ell' +2$. 
Hence, the new block $B_{\ell+1}$ is formed by the left block plus a single site. 

\vspace{5mm}

\begin{figure}[hbt]
\begin{center}
  \epsfig{file=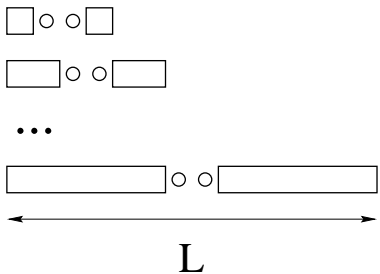,width=5cm}
\end{center}
\caption{Schematic infinite system algorithm.}
\label{fig_infinite}
\end{figure}

In general, DMRG algorithms can be divided into two classes according to how 
the environment block $B_{\ell'}$ is chosen. In the infinite system algorithm,  
$B_{\ell'}$ is chosen to be  $B^R_\ell$, the reflection of  $B_\ell$. 
Therefore the superblock grows as  $B_{\ell}$ is built up. 
For the finite system algorithm, $B_{\ell'}$ is usually chosen so 
that the length of the superblock is $L$.
It begins with the use of the infinite system algorithm (see Fig. 
\ref{fig_infinite}) for $\frac{L}{2}-1$ steps, such that the 
final superblock obtained is of size $L$. 
That is, we start with a four site chain $B_1 \bullet \bullet B_1^R$. 	
Then we calculate the density matrix and construct an effective Hamiltonian for $B_2$. 
In the second step, we diagonalize $B_2 \bullet \bullet B_2^R$ where we have
formed $B_2^R$ by reflecting $B_2$. 
We continue in this manner, diagonalizing
the configuration $B_\ell \bullet \bullet B_\ell^R$, setting $B'_{\ell+1}$ equal
to $B_\ell \bullet$, and then using $B_{\ell+1}$ and its reflection 
in the next step. 
Once the system  $B_{\frac{L}{2}-1} \bullet \bullet B_{\frac{L}{2}-1}^R$ is 
used to form $B_{\frac{L}{2}}$, the next step is to form $B_{\frac{L}{2}+1}$.
This system and all the other superblocks to follow will involve exactly
$L$ sites.
We continue to form the other blocks up to size $L-3$ using the superblock 
$B_\ell \bullet \bullet B^R_{L-\ell-2}$ to build $B_{\ell+1}$.
This sequence of steps corresponds to the first iteration  of the finite system 
algorithm.

\begin{figure}[hbt]
\begin{center}
  \epsfig{file=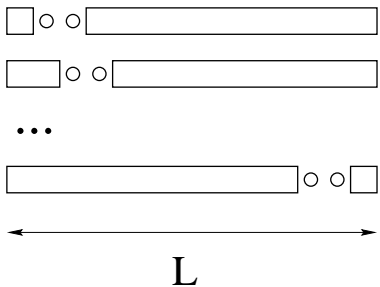,width=5cm}
\end{center}
\caption{Schematic finite system algorithm}
\label{fig_finite}
\end{figure}
The second and subsequent iterations  use superblocks of 
fixed length $L$ (see Fig. \ref{fig_finite}). 
We start by diagonalizing $B_1 \bullet \bullet B_{L-3}^R$ forming $B_2$ up 
to $B_{L-3} \bullet \bullet B_1^R$ where we use
the blocks obtained from the previous iteration to build the reflected blocks
$B_\ell^R$.
On the very last iteration we stop after diagonalizing $B_{\frac{L}{2}-1} 
\bullet \bullet B_{\frac{L}{2}-1}^R$ and then use the wave function of the 
$L$ sites system to measure ground-state properties, 
e.g. correlation functions. 
At each sweep (left-to-right and right-to-left iteration) we increase the number 
of states $m$ used to represent the Hamiltonian and other operators inside the
block. 

Since we truncate the basis, the algorithm is variational and in practice the 
calculated value for the ground-state energy obeys the empirical law
\begin{equation}
     E (m) = E_0 + \Delta e^{-m/m^*}
\label{eq:empeq}
\end{equation}
where $E_0$ is the targeted energy;
$\Delta$ and $m^*$ are adjustable parameters.

An important improvement in efficiency can be made by keeping track of the
wave function in the process of adding a site \cite{White96}. We can then 
construct a better initial wave function for the Davidson 
diagonalization of the superblock, thus reducing the number of steps needed to 
obtain convergence.


\section{Quantum Chemistry with the DMRG}

\subsection{The non-relativistic time-independent Schr\"odinger equation}

We will now show how the DMRG can be applied to ab initio quantum chemistry
\cite{WhiteMartin}. 
Essentially the DMRG will yield both the energy and the wave function of the ground
state and of some excited states. 
More precisely, we consider a molecule with $N$ interacting electrons moving in the field of $M$ fixed nuclei.
The Hamiltonian describing this system is 
\begin{equation}
 H = -\frac{1}{2} \sum_{i=1}^N   \Delta_i
  -   \sum_{i=1}^N \sum_{k=1}^M \frac{Z_ke^2}
    {|\vec{r_i} -\vec{r_k} |}
  + \frac{1}{2}  \sum_{i,j} \frac{e^2}
    {|\vec{r_i} -\vec{r_j} |}  
\label{eq:ham1q}
\end{equation}
and we will use atomic units throughout this paper.

\subsection{Second quantized form of the Hamiltonian}

To perform calculations, this Hamiltonian is represented in a
finite basis of $K$ functions.  One chooses usually atomic
orbitals $ \chi_i  $ with $i=1, ... , K$ which are described,
for example, by the product of a few Gaussians and spherical
harmonics.  This set expands a smaller (finite size) Hilbert
space and is not orthogonal.  The first step is to orthogonalize
these functions to be able to represent the Hamiltonian
(\ref{eq:ham1q}) in second quantization.  In practice one can
perform a Hartree-Fock calculation in order to generate a set of
basis functions.  In doing so, we have a good (low energy)
starting point for the calculation of the correlation energy.

Consider a set of $K$ molecular orbitals (MO), $\{\varphi_i$\},
$i=1, ... , K$, which we reorder with increasing energy.  We
mainly used Hartree-Fock molecular orbitals, but we do not have
to restrict ourselves to them, and we have also used Kohn-Sham
orbitals (see discussion in \ref{DFTorbital}).  
In case of closed shell systems, an approximate
ground state would be when the first $N/2$ orbitals are occupied
by 2 electrons and the remainders (virtual orbitals) are empty.
These MO can be obtained from the atomic orbitals by the
well-known Linear Combination of Atomic Orbitals - Self
Consistent Field (LCAO-SCF) scheme\cite{LCAO-SCF}. One uses
then the second quantization formalism to reexpress the
Hamiltonian (\ref{eq:ham1q}).  An introduction to that technique
can be found, e.g., in Ref.  \onlinecite{OstlundSzabo}.  The
creation (annihilation) operator $c_{i\sigma}^\dag$
($c_{i\sigma}$) creates (annihilates) an electron with spin
$\sigma$ in the MO $\varphi_i$.  The Hamiltonian we shall use
from now on reads then
\be
    H = \sum_{i,j} \sum_\sigma H_{ij} c_{i\sigma}^\dag c_{j\sigma}
 + \frac{1}{2} \sum_{i,j,k,l} G_{ijkl} \sum_{\sigma , \sigma'}
  c_{i\sigma}^\dag c_{j\sigma'}^\dag c_{k\sigma'} c_{l\sigma} .
\label{eq:ham2q}
\ee
The sum is over all orbitals and spin $\sigma$. 
The integrals
\be
    H_{ij} = \int d\vec{r} \varphi_i^* (\vec{r})  \left[ -\frac{1}{2}\Delta
- \sum_{k=1}^M  \frac{Z_k e^2}{|\vec{r} -\vec{r_k} |}
\right]    \varphi_j (\vec{r})
\ee
represent the elements of the core Hamiltonian (kinetic energy
and nuclear attraction), while
\be
    G_{ijkl} =  \int \int d\vec{r}  d\vec{r'}  \varphi_i^* (\vec{r})
   \varphi_j^* (\vec{r'})  \frac{1}{|\vec{r}-\vec{r'}|}
 \varphi_k (\vec{r'}) \varphi_l (\vec{r})
\ee
are the interelectronic repulsion integrals.  Using the
occupation number formalism, a Slater determinant $\psi$ of $N$
electrons can be written as
\be
    \psi = c^\dag_{i_1 \sigma_1}  c^\dag_{i_2 \sigma_2} \ldots |0\rangle 
\ee
where $ |0\rangle $ is the vacuum state.
As an example, the singlet Hartree-Fock eigenstate for a closed shell
system is
\be
   \psi^0_{\mbox{\footnotesize HF}} = c^\dag_{1\up}c^\dag_{1\dn}
c^\dag_{2\up}c^\dag_{2\dn}  \ldots c^\dag_{N/2 \up} c^\dag_{N/2 \dn} |0\rangle .
\ee 
A full-CI calculation is then the exact diagonalization of
(\ref{eq:ham2q}) in this occupation number representation of the
Hilbert space basis.  Since the size of the matrix of
(\ref{eq:ham2q}) is ${\cal O} (N!)$, e.g. 10 spin-up electrons
and 10 spin-down electrons with 40 orbitals has a dimension of
approximatively $10^{18}$, one cannot use the
Lanczos\cite{Lanczos} or Davidson\cite{Davidson} algorithm to
solve this problem.  Instead we use the DMRG to obtain the
ground state and some excited states of the Hamiltonian
(\ref{eq:ham2q}) for small to medium sized polyatomic molecules.

Since the DMRG was initially developed for one-dimensional quantum lattices,  a 
molecular wave function (such as the Hartree-Fock MO) becomes a site which 
can be empty,
singly occupied with an up or down-spin electron or doubly occupied.
Hence a molecule described by $K$ orbitals becomes a lattice with $L=K$ sites
and the Hamiltonian given in Eq. (\ref{eq:ham2q}).
As input for the DMRG we need the  matrix elements $H_{ij}$ and $G_{ijkl}$. 
Since our Hamiltonian conserves the total number of particles and the total spin,
we also specify the number of electrons $N$ and the projection of the total spin
on a quantization axis $S_z$ in order to reduce the dimension of the Hilbert space.

\subsection{One DMRG step}

We want to describe here explicitly the major part of the DMRG applied to 
quantum chemistry,  a renormalization step with the density matrix.
We use the finite-size algorithm of the DMRG and target only the ground state.
In a warmup procedure, we construct an approximated ground state 
(e.g. Hartree-Fock). 
Then we start the algorithm following the description in Ref. 
\onlinecite{WhiteNoack98}.
Let us consider in detail the addition of an arbitrary orbital (site).
We already have built a subspace $S_{\ell} $ which describes partially the electronic
structure of the molecule using the first $\ell$ MOs. 
In the original DMRG formalism this subspace spanned by $\ell$ basis functions, which
were Wannier functions located on each site, was called a block.
To avoid possible confusion with quantum lattice concepts, we prefer here to use
the word ``subspace".
For the description of that subspace we keep only $m$ states.
These states are used to represent i) the part of the ground state in that 
subspace; ii) the internal part of the Hamiltonian, i.e. terms that only 
involve indices which belong to the subspace ($i,j,k,l \leq \ell$);
and iii) terms that will be used in constructing the interactions
between the subspaces, such as
$ \left[ c^\dag_{i\sigma} \right], \left[ c_{i\sigma} \right],
	 \left[ c^\dag_{i\sigma} c^\dag_{j\sigma} \right], \ldots$. 

\begin{figure}[hbt]
\begin{center}
  \epsfig{file=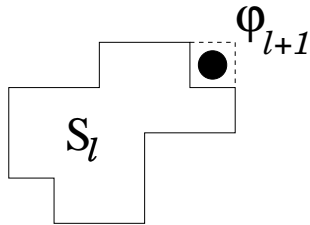,width=4cm}
\end{center}
\caption{Schematic view of the left subspace with one orbital added.}
\label{fig:subspace1}
\end{figure}

\begin{figure}[hbt]
\begin{center}
  \epsfig{file=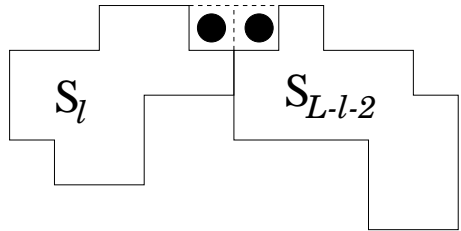,width=6cm}
\end{center}
\caption{Schematic view of both subspaces.}
\label{fig:subspace2}
\end{figure}

We add then the $(\ell + 1)^{\mbox{\footnotesize th}}$ orbital (site) 
to the current subspace (see Fig. \ref{fig:subspace1})
and construct the internal Hamiltonian for the new subspace $S_{\ell + 1}$
and for all other operators needed as well. 
\vspace{5mm}
\noindent
The  Hamiltonian reads
\begin{eqnarray}
    H_{\ell + 1} &=& H_{\ell} \otimes {\bf 1} +  {\bf 1} \otimes H_1
    + \sum_{i \leq \ell} \sum_\sigma H_{i,\ell+1}  \left[ c^\dag_{i\sigma} \right]_\ell
    \otimes \left[ c_{\ell+1,\sigma} \right]_1 + \ldots \\
   & + &\sum_{i,j \leq \ell + 1} \sum_{ \sigma, \sigma'} G_{ij,\ell + 1, \ell + 1} 
  \left[ c^\dag_{i\sigma} c^\dag_{j\sigma'} \right]_\ell \otimes
  \left[ c_{\ell+1,\sigma'} c_{\ell+1,\sigma} \right]_1
  + \ldots
\end{eqnarray}
where $[A]_\ell$ denotes the matrix of the operator $A$ represented in the basis of 
the $m$ most significant states of the subspace $S_\ell$ and $[A]_1$ denotes the 
matrix of the operator $A$ represented in the Hilbert space of one orbital.
For example, if the order of the basis set is 
 $\left\{ | 0 \rangle, | \uparrow \rangle, | \downarrow \rangle 
, | \uparrow \downarrow \rangle \right\}$, where 
$| \uparrow \downarrow \rangle $
is defined as $c^\dag_\uparrow c^\dag_\downarrow | 0 \rangle$,
the matrix for the creation operator
of an up-spin electron for the MO $\ell + 1$ is
\be
 \left[ c^\dag_{\ell+1,\up}  \right]_1 = 
  \left( \begin{array}{cccc}
    0 & 0 & 0 & 0 \\
    1 & 0 & 0 & 0 \\
    0 & 0 & 0 & 0 \\
    0 & 0 & 1 & 0 
  \end{array} \right).
\ee
The new subspace is now described by $4m$ states.
We will illustrate now how to reduce this number to $m$. 
To this purpose we need to construct the environment which involves all the 
residual orbitals. 
In case of a molecule, we can not reflect the current subspace
(block) as in case of a quantum lattice system since there is,
in general, no equivalent reflection symmetry.  We apply the
same procedure for the right subspace $S_{L-\ell-2}$.
In Fig. \ref{fig:subspace2} we show the construction of the left and right
subspaces.
\noindent
At the end we will have two subspaces $S_{\ell+1}$ and
$S_{L-\ell-1}$; the first describing  partially the electronic
structure of the molecule within the first $\ell + 1$ orbitals
and the second one, the rest of the molecule with the remaining
$L-\ell-1$ orbitals.  In order to retain only the $m$ states of
$S_{\ell +1}$ which best represents the whole molecule, we
construct the full molecule by adding the two subspaces together.
We build a representation of the full Hilbert space by adding the two
subspaces together, which means taking the direct product of the
two subspaces. We then compute the  ground 
state of the whole molecule using the Davidson algorithm.  \cite{Davidson} 
We then construct the projected density matrix.  
The full Hamiltonian could be constructed in the
same way as when adding two blocks together, Eq. (19), 
but to be more efficient, we use the
following procedure.  In the Davidson algorithm, the time
consuming part is the $H |\Psi\rangle $ multiplication.  The
matrix of the molecular Hamiltonian can be written in the
general form as
\be
   \left[ H^{\mbox{\footnotesize molecule}} \right]_{ij;i'j'} = 
	\sum_\alpha A^\alpha_{ii'} B^\alpha_{jj'}
\ee
where the index $i$ and $i'$ stands for states in the left
subspace and $j,j'$ for the right subspace.  The sum on $\alpha$
is over all terms of the Hamiltonian.  Then the product $
H^{\mbox{\footnotesize molecule}} |\Psi\rangle $ can be written
as 
\be
    \sum_{i',j'}  \left[ H^{\mbox{\footnotesize molecule}} \right]_{ij;i'j'} 
     \Psi_{ij} = \sum_\alpha \left[ \sum_{i'} A^\alpha_{ii'}
    \left( \sum_{j'} B^\alpha_{jj'} \Psi_{i'j'}
    \right) \right]
\ee
The last sum is performed first as a matrix multiplication of
$B^\alpha$ with $\Psi^T$, to form a temporary matrix
$C^\alpha_{ji'}$.  Then a matrix multiplication of $A^\alpha$ by
$[C^\alpha]^T$ yields a partial term, which is added into the
resulting vector, giving a sum on $\alpha$.  Using the Davidson
algorithm one obtains the ground state $\Psi_0$ and some excited
states, if desired.

Then we construct the reduced density matrix tracing out the
environment components 
\be
    \rho_{ii'} = \sum_{j} \Psi^*_{ij}  \Psi_{i'j} .
\ee
One diagonalizes $\rho$, obtaining $4m$ eigenvectors $\phi^\nu$ and the
corresponding eigenvalues $\omega_\nu$. 
Each $\omega_\nu$ represents the probability of the system being in the
state $\phi^\nu$, with
\be
    \sum_{\nu=1}^{4m} \omega_\nu = 1 .
\ee
One chooses the $m$ states with the highest $\omega_\nu$ value.
The discarded weight
\be
     1 -  \sum_{\nu=1}^{m} \omega_\nu 
\ee
measures the accuracy of the truncation to $m$ states.

The reader may notice a similarity between this procedure and the
construction of natural orbitals. However, a crucial difference
is that the density matrix here is a many particle density
matrix, with an exponentially large dimension, not a single
particle density matrix. The resulting basis is a many particle
basis, with each state representing a complicated sum over large
numbers of configurations. 

When the new basis is formed ($m$ elements), the operators must be updated.
The eigenstates  $\phi^\nu$ can be written in the form
\be
     \phi^\nu_{ij}
\ee
where $i$ denotes a state in the subspace $S_\ell$ and state $j$ corresponding
to the $(\ell + 1)^{\mbox{\footnotesize th}}$ orbital.
We store them in a $m \times 4m$ matrix ${\cal O}$.
Thus each operator $A$ that is needed is replaced by 
\be
    \tilde{A} = {\cal O} A {\cal O}^T .
\ee
Adding the $(\ell + 1)^{\mbox{\footnotesize th}}$ orbital to the subspace
$S_\ell$ corresponds to  one DMRG step. 
There are a total $(K-2)$ DMRG steps to zip through all $K$ orbitals,
and another $(K-2)$ DMRG steps to zip back to the beginning,
with the roles of the left and right subspaces switched.
This is a complete DMRG sweep, keeping always $m$ states per RG step.
After one iteration we increase the number of significant states  $m$, e.g. by 
doubling it, and start again using the preceding result for 
$S_{L-\ell-2}$ to describe environment (i.e. the complementary space).
Since we truncate the basis of the full Hilbert space but
calculate the Hamiltonian exactly within this basis,
this algorithm is variational. 
In the limit where we keep enough states $m$ to solve the 
full-CI problem, this algorithm is also clearly size consistent.

At the end, we obtain the energy and a representation of the ground state
from which we can compute the density, the natural orbitals, 
the two-electron density matrix, etc.

\subsection{Illustration}

\begin{figure}[hbt]
\begin{center}
  \epsfig{file=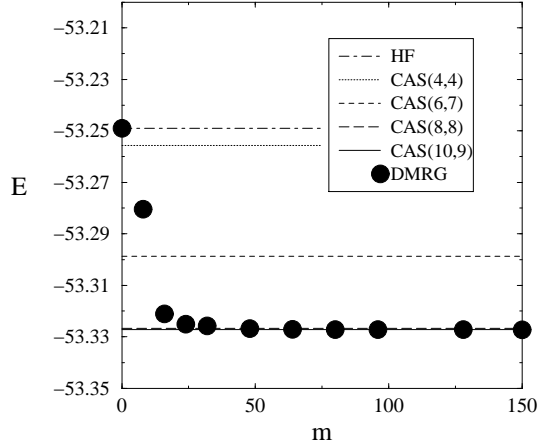,width=7cm}
\end{center}
\caption{Ground-state energy $E$ of CH$_4$ as a function of the number of
states retained $m$ at each DMRG iteration. 
The lines are the Hartree-Fock and CAS results.}
\label{fig:ch4}
\end{figure}

In order to illustrate the DMRG procedure, let us consider a simple
example, the methane molecule with $T_d$ symmetry (CH$_4$). 
We follow the procedure outlined above :
\begin{enumerate}
\item Choose  a minimal basis  : STO-3G (9 orbitals).
\item Perform a Hartree-Fock calculation obtaining 9 MO and the Hartree-Fock
electronic energy $E = -53.248968 \ldots$
\item Represent the Hamiltonian in this basis, i.e. calculate all $H_{ij}$ and $G_{ijkl}$.
\item Apply the DMRG to obtain the ground state of the molecule 
(see Appendix for some technical points).
\end{enumerate}
In Fig. \ref{fig:ch4} we show the ground-state energy $E$ as a function of 
the number of significant states $m$ retained in each DMRG iteration.
We also show results obtained by the CAS method. 
In CAS($\nu$,$\mu$) we consider the complete active space configuration
interaction of $\mu$ electrons for $\nu$ orbitals. 
We see that DMRG is variational and converges to the CAS(10,9) result which
is full CI.

In Table \ref{table:occ} we also show the occupation numbers
\be
    \langle n_i \rangle = \langle c^\dag_{i\up} c_{i\up} \rangle 
           + \langle c^\dag_{i\dn} c_{i\dn} \rangle 
\ee
of the MO as a function of the progress of calculation.
For instance it can be seen that the occupation of
symmetry equivalent orbitals ($t_{2_i}$) becomes equal at the last iteration.
DMRG achieves the accuracy of CAS(8,8) (occupation of $1a_1$ orbital frozen) 
when $m>32$. 
The analysis of the occupation numbers shows the importance of the 
higher energy states in the description of the
ground state of this typical closed-shell system. 
A significant occupation of the orbitals usually unoccupied in the HF
procedure is observed. 

\begin{table}
\[
\begin{array}{|c|c|c|c|c|c|}
\hline
\mbox{Irreducible} & \mbox{Hartree Fock} & \langle n_i \rangle \mbox{ after} 
& \langle n_i \rangle \mbox{ after} &  &  \langle n_i \rangle  \mbox{ after} \\ 
\mbox{representation} & \mbox{eigenvalue} & \mbox{warmup} & \mbox{first iteration} 
& \ldots & \mbox{last iteration} \\ \hline  
1a_1      &  -11.02969 & 2.000000  & 1.999997 & \ldots  & 1.999965  \\
2a_1      & -0.91210 & 2.000000 &   1.998077& \ldots    & 1.985416  \\
1t_{2_1}  & -0.52050 & 2.000000 &  1.996863 & \ldots	& 1.976450  \\
1t_{2_2}  & -0.52050 & 2.000000 &  1.992911 & \ldots	&   1.976450  \\ 
1t_{2_3}  & -0.52050 & 2.000000 &   1.988980 & \ldots	&   1.976450  \\
2t_{2_1}  & 0.71922 & 0.000000 &   0.021086  & \ldots	&   0.021952  \\
2t_{2_2}  & 0.71922 & 0.000000 &   0.005932  & \ldots	&    0.021952  \\ 
2t_{2_3}  & 0.71922 & 0.000000 &   0.010275  & \ldots	& 0.021951  \\
3a_1     & 0.76106 & 0.000000 &    0.009580 & \ldots	 &   0.019412 \\ \hline
\end{array}
\]
\caption{Occupation numbers $\langle n_i \rangle$ of the MO as a function
of the progress of the calculation.}
\label{table:occ}
\end{table}


\section{Results and discussion}

In order to demonstrate the performance of the DMRG when applied to quantum 
chemical  problems, we have chosen a simple
magnetic system which has already been the subject of extensive post-HF and DFT 
studies. 
In fact, computing the energy difference between multiplets of different 
spins is a  difficult task for standard
computational approaches. 
This is due to the very high accuracy required for the 
calculation of the ground- and excited-state energies and the intrinsic need 
of taking into account the mixing with states of
higher energy (configuration interaction).
For this reason, we believe that this calculation can be a useful benchmark 
to test the performance of DMRG in quantum chemistry.
The crucial features of any new quantum chemical method are scaling and 
convergence.
In the following we address this point,
both when starting from localized and non-localized Hartree-Fock MO 
as well as from Kohn-Sham orbitals. 

\subsection{HHeH}

\begin{figure} [htb]
\begin{center}
\begin{picture}(100,30)
\put(10,0){\circle*{20}}
\put(10,0){\line(1,0){30}}
\put(55,0){\circle{30}}
\put(70,0){\line(1,0){30}}
\put(100,0){\circle*{20}}
\end{picture}
\end{center}
\vspace{10mm}
\caption{HHeH: A schematic sketch}
\label{p:hhehstr}
\end{figure}
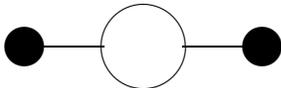

The linear HHeH molecule (Fig. \ref{p:hhehstr}) was chosen as a typical example 
for a model magnetic system. In this simple case the two paramagnetic
centers, carrying each an unpaired spin, are the hydrogen atoms. 
They are linked via a diamagnetic bridge constituted by the He 
atom. The spins of the two paramagnetic electrons can be parallel or 
antiparallel yielding two different spin states, namely a singlet ($S=0$) 
and a triplet ($S=1$).
The difference in energy between these two spin states as a function of the 
H-He distance, has been subject of both
post-HF and DFT calculations \cite{HHeHpa,Rappe}, and shows the importance
of electron correlation.

In Table II the results found in literature \cite{HHeHpa} are 
compared  to the exact results obtained with both
full-CI approach and DMRG using the same basis set (6-311G** \cite{Pople}). 
This basis set contains a total of 18 orbitals. 
The following notation is used: 
CAS($\nu$,$\mu$) = complete active space configuration interaction 
considering $\nu$ electrons and $\mu$ orbitals; 
QCISD(T) = quadratic CI including single and double excitations and also triple 
excitations perturbatively; 
BS = broken symmetry \cite{BS}; 
SD = single determinant \cite{SD}; 
LDA = local density approximation for exchange correlation functional in the 
Vosko-Wilk-Nusair parameterization \cite{VWN}; 
BP = generalized gradient approximation for exchange and correlation 
functional using the Becke parameterization \cite{Becke}.
The data obtained with post-HF methods show that a large part of electron 
correlation is already included  at a modest CAS level.
Both DF based techniques, the single determinant method 
\cite{SD} (SD) and the broken symmetry approach \cite{BS} (BS), 
which are less expensive in terms of computational 
resources, are just adequate to yield qualitative results. 

\begin{table}[htbp]
\begin{center}
\[
\begin{array}{|c|c|}
\hline
\mbox{Method} & \mbox{Distance=1.25 $\AA$}  \\
\hline
\mbox{CAS(2,2)}&  4204  \\
\mbox{CAS(4,3)}&  4294  \\
\mbox{QCISD(T)}&  4298   \\
\mbox{BS-LDA}&  12432  \\
\mbox{BS-BP} &  10529  \\
\mbox{SD-LDA}  &  9050  \\
\mbox{SD-BP} &  7799   \\
\mbox{Full-CI} &  4860 \\
\mbox{DMRG} &  4859  \\
\hline
\end{array}
\]
\caption{HHeH Singlet-triplet energy gap (in cm$^{-1}$) as a function of the 
H-He distance for various methods. }
\end{center}
\label{table:hheh}
\end{table}

\begin{figure} [htbp]
\begin{center}
\epsfig{file=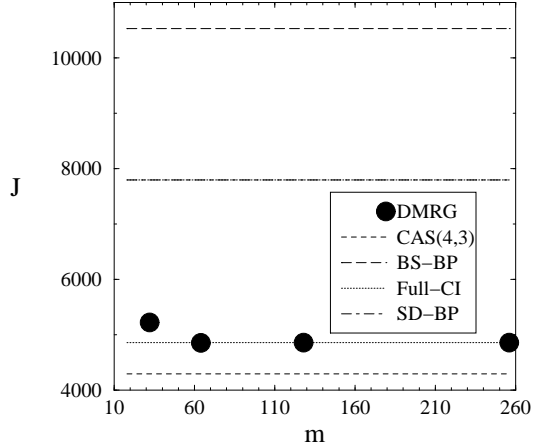,width=7cm}
\end{center}
\caption{Singlet-triplet energy gap of HHeH as a function of $m$ for a H-He distance of 
1.25 $\AA$}
\label{p:j}
\end{figure}

In Fig. \ref{p:j} the singlet-triplet energy gap as 
a function of $m$ is reported for a H-He distance of 1.25 $\AA$.
For $m>64$ the DMRG procedure has converged to the 
exact result (full CI). 
Results superior to those obtained with DFT are observed
already when $m \approx 32$. 
The same level of accuracy as with CAS(4,3) is also reached for this $m$ value.

\subsection{Choice of the basis}
\label{DFTorbital}

\begin{figure} [ht]
\begin{center}
\epsfig{file=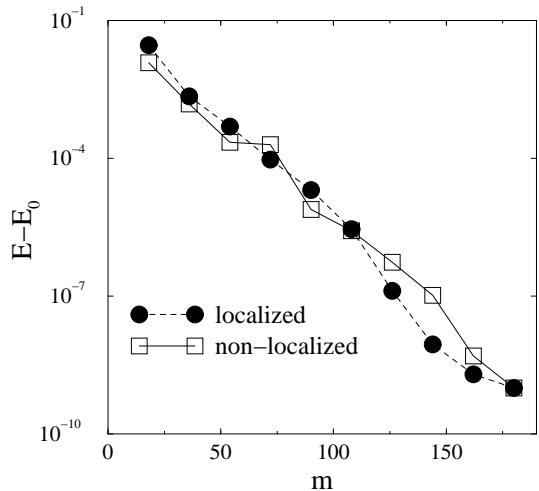,width=7cm}
\end{center}
\caption{Rescaled energy $E-E_0$ as a function of $m$ using localized 
and non localized HFMO on a lin-log scale. The dashed lines are just guides
for the eyes. } 
\label{p:loc}
\end{figure}

The results obtained with DMRG when applied to quantum lattice systems 
\cite{DMRGbook}  suggests that the use of localized basis sets, which 
reduce the 
interactions between far distant neighbors, will enhance convergence. 
Therefore we attempted to localize the starting MO basis set
using standard localization techniques.
Amongst the classical quantum chemical localization techniques 
\cite{Boys,ER,vN}, we have chosen to apply the procedure of Mezey and Pipek 
\cite{Pipek} due to the fact that just overlap integrals are
needed to perform the localization.
This procedure is based on the minimization of a mean localization measure 
of the occupied molecular orbitals $D$ defined as
\be 
    D^{-1} = N^{-1} \sum_{i=1}^N d_{i}^{-1}
\ee
with
\be 
    d_{i}^{-1} = \sum_{k=1}^M (Q^{i}_{k})^{2}
\ee
and where $k$ are the different nuclei of the molecule, $i$ the molecular orbitals, 
and $Q^i_k$ are the gross atomic Mulliken population of the orbital $i$.
Rather than localizing all the MO, we only localized
the subset of the virtual MO (not occupied at a HF level).
The two subsets, formed by the occupied non-localized MO
and the virtual localized MO, are still orthonormal. 
In fact, the localization of the whole set of MO would yield a substantial 
increase of the starting energy of the DMRG procedure, while localizing the 
virtual orbitals only does not. 
Recalling the empirical dependence of the electronic energy upon $m$ 
(Eq. \ref{eq:empeq}), we show in Fig. \ref{p:loc}, $E(m) - E_0$ versus $m$ to
illustrate the convergence of the DRMG procedure 
using localized and non-localized Hartree-Fock MO.
Unfortunately, we observe that the localization procedure does not yield
a faster convergence. 
In fact, it should be stressed that
this system is not the most adequate to test the impact of localization.
Indeed, there are only a few electrons involved and the spatial extent of the
molecule is rather small.
Nevertheless, we expect that this localization procedure
should improve the convergence of the DMRG for larger systems, e.g. linear
chain like molecules for which most of the $V_{ijkl}$ terms will vanish or
become negligible when using localized basis functions.

\begin{figure} [ht]
\begin{center}
\epsfig{file=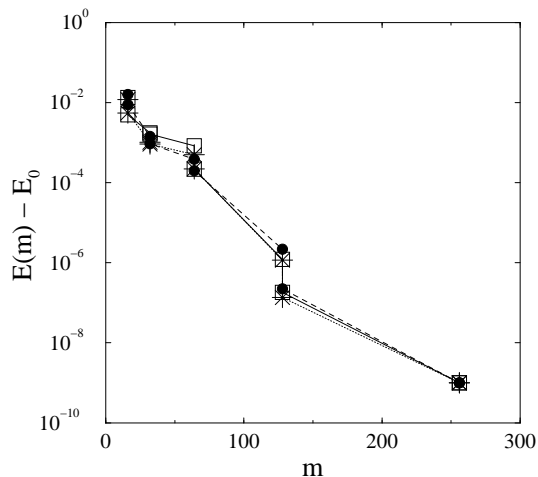,width=7cm}
\end{center}
\caption{Rescaled energy $E-E_0$ as a function of $m$ using HF, LDA and GGA 
orbitals, on a lin-log scale.}
\label{p:rks}
\end{figure}

Another feature we investigated was the possibility to use Kohn-Sham orbitals
as basis functions. 
We performed DMRG calculation starting from Hartree-Fock orbitals as well as 
from Kohn-Sham orbitals obtained using both LDA \cite{VWN} and GGA\cite{PW91} 
functionals.
The results obtained for the singlet state with a H-He distance of 1.25 $\AA$ 
are shown in Fig. \ref{p:rks}.
We see that all three basis functions have essentially the same 
convergence rate. 

\subsection{Scaling of the method}

In order to test the numerical effort spent with DMRG we performed several 
tests on the HHeH singlet state at a distance of 1.25 $\AA$, either using 
different basis sets (varying $K$ at constant $m$) or keeping the 
basis set (varying $m$ at constant $K$). 
The aim of these calculations was twofold. 
First, to test the scaling of DMRG \cite{WhiteMartin} and second, to 
demonstrate its superiority with respect to standard post-HF approaches
(full CI, QCISD or QCISD(T)). 
The main disadvantage of post-HF solutions (apart from their scaling) 
is the lack of at least one of the main feature of DMRG: size consistency and 
variational character.
The only theoretical approach which fulfills all these requirements is 
presently full CI.
For example, CID or QCISD methods are variational but not size consistent.
On the other hand, all perturbative techniques (MPn), 
whose scaling are comparable with the 
one of DMRG, are not variational and practically feasible only up to  
$n=4$ (inclusion of quadruple excitations). 

\begin{figure} [ht]
\begin{center}
\epsfig{file=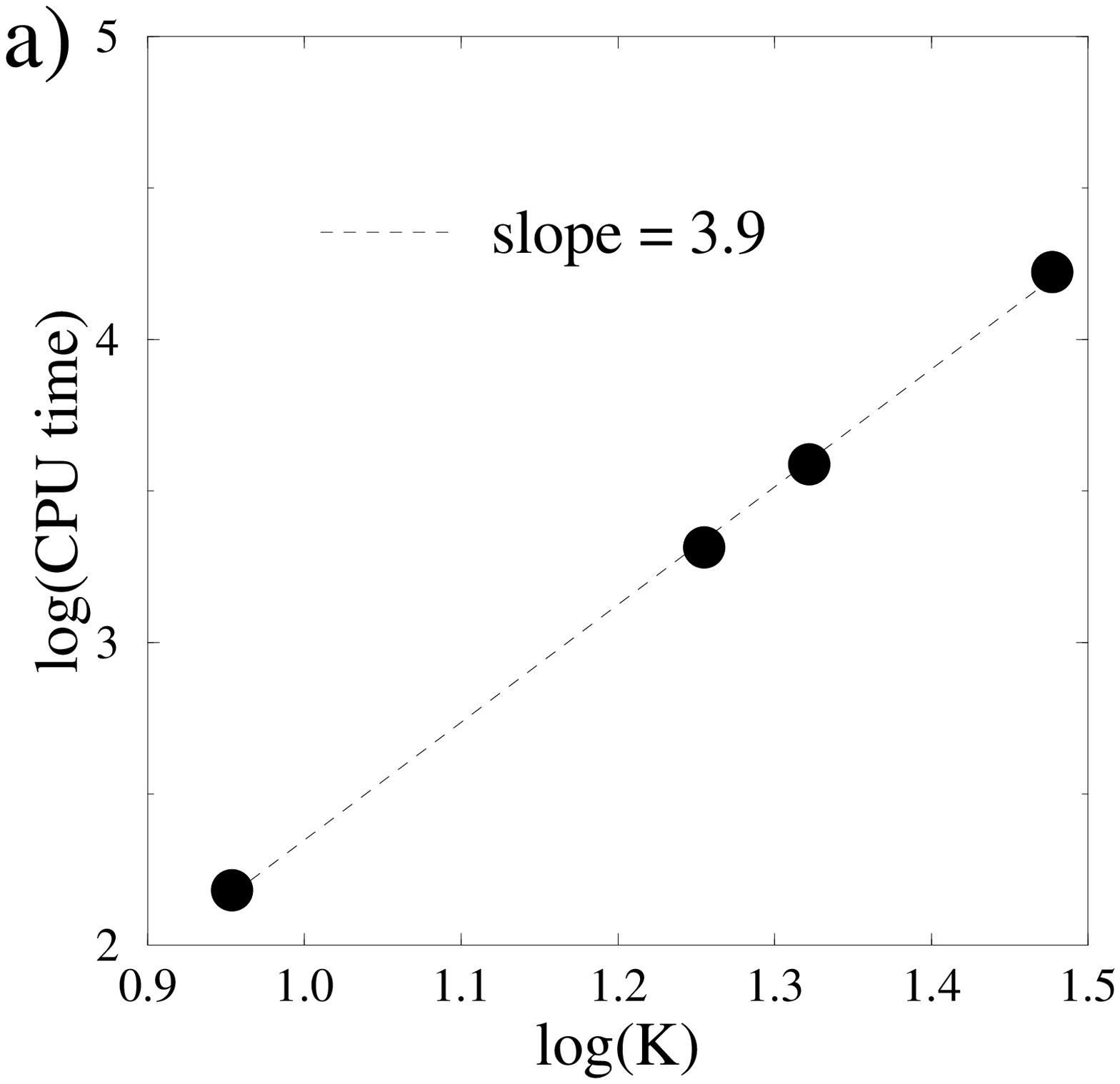,width=6cm}
\epsfig{file=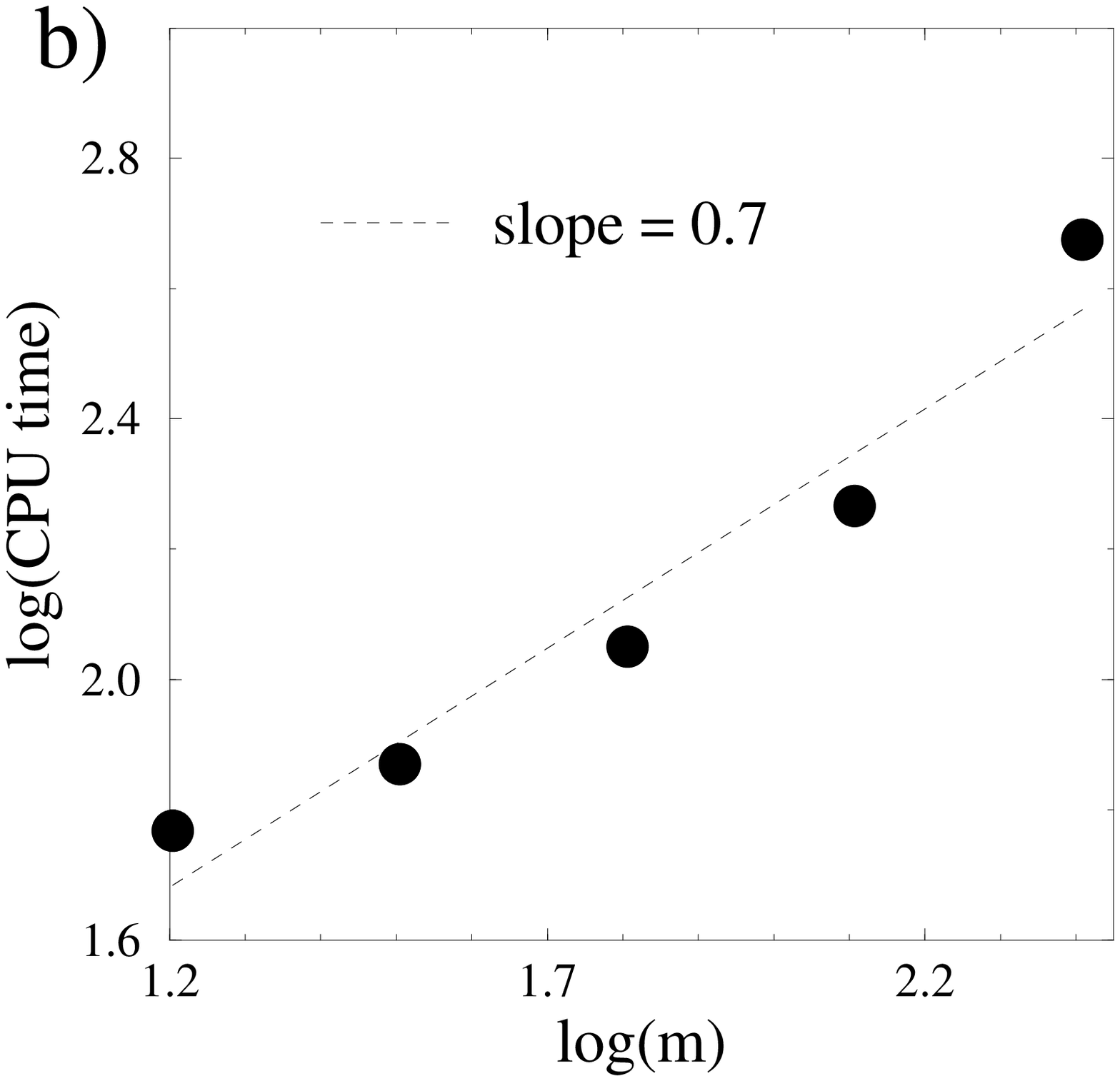,width=6cm}
\end{center}
\caption{ a) CPU time as a function of $K$ on a log-log scale;
b) CPU time as a function of $m$ on a log-log scale. 
Both plots are for HHeH.}
\label{p:sclN}
\end{figure}

The theoretical scaling of DMRG (neglecting the cost of the preliminary 
Hartree-Fock procedure) is ${\cal O} (K^4m^2)$.
In our case $K$ is relatively small ($K_{\mbox{\footnotesize max}} = 30$),
while $m$ can be 
an order of magnitude bigger than $K$ ($m_{\mbox{\footnotesize max}} = 264$).
In Fig. \ref{p:sclN} the plots of the CPU time versus $K$ at fixed $m$ 
and versus $m$ at fixed $K$ are shown. 
The slope, obtained by linear regression,
 observed in part a) is in  agreement with the theoretical 
predictions, i.e. ${\cal O}(K^4)$.
While the slope in part b) is lower that predicted since we use highly
optimized BLAS to perform the matrix operations (cf Appendix).
Thus the scaling of DMRG appears to be competitive even with 
QCISD, which scales as ${\cal O}(K^7)$, while yielding much better results.

\section{Acknowledgments}

IC, CD and SD acknowledge support from the Swiss National Science Foundation,
SD also acknowledges support from the ITP under NSF grant \#PHY94-07194
and SRW acknowledges support from the NSF under grant \#DMR98-70930.

\appendix

\section{Implementation of the method}
\label{appendix}

The program we use is written in C++, an object-oriented language. 
We have to keep track of all sorts of matrices connecting different subspaces
with different quantum numbers,
which is conveniently done using classes and overloading operators.
In addition, the program uses a matrix library, developed by S. White 
and R. Noack, which links the matrix-matrix and matrix-vector multiplication 
to corresponding BLAS to be more efficient.
In order to save memory, only the information for the specific blocks in use
are loaded, while the rest is stored on disk.
For the moment we cannot exceed a number of basis function of $K=40$
(which uses about 2 Gb of memory).
And right now we are working on rewriting the program in order to be able
to handle 100 basis functions.

The program runs under Unix and has mainly be used on workstations such as 
Linux PC, Dec Alpha, HP, SGI and Cray.
A tutorial program for a particle in a box and the matrix library can be found
at http://hedrock.ps.uci.edu/.


\end{document}